\documentclass[a4paper,review,11pt,authoryear]{elsarticle}
\journal{arXiv}

\usepackage{natbib}

\usepackage[margin=1in]{geometry}
\usepackage{array}
\usepackage{booktabs}
\usepackage{bm}
\usepackage{amsthm}
\usepackage{arydshln}
\usepackage{amsmath}
\usepackage{amstext}
\usepackage{mathtools}
\usepackage{amsfonts}
\usepackage[font=small]{caption}
\usepackage{graphicx}
\usepackage{natbib}
\usepackage{lineno}
\usepackage{bbm}
\usepackage{xurl}
\usepackage{verbatim}
\usepackage{setspace}
\usepackage{multirow}
\usepackage{hyperref}
\hypersetup{
  colorlinks=true,
  linkcolor=blue,
  citecolor=blue,
  urlcolor=blue}

\usepackage{verbatim}
\newcommand{\pkg}[1]{{\normalfont\fontseries{b}\selectfont #1}}
\let\proglang=\textsf
\let\code=\texttt

\usepackage{todonotes}

\usepackage{enumitem}
\bibpunct{(}{)}{;}{a}{,}{,}
\makeatletter
\renewcommand{\paragraph}{\@startsection{paragraph}{4}{0ex}%
   {-3.25ex plus -1ex minus -0.2ex}%
   {1.5ex plus 0.2ex}%
   {\normalfont\normalsize\bfseries}}
\makeatother

\graphicspath{{figures/}}

\newtheorem{thm}{Theorem}

\begin{document}
\begin{frontmatter}

\title{Optimal reconciliation with immutable forecasts}

\author[label1]{Bohan Zhang}
\address[label1]{School of Economics and Management, Beihang University, Beijing, China}

\author[label1]{Yanfei Kang}
\author[label2]{Anastasios Panagiotelis}

\author[label3]{Feng Li\corref{cor1}}
\ead{feng.li@cufe.edu.cn}
\cortext[cor1]{Correspondance: Feng Li, School of Statistics and Mathematics, Central University of Finance and Economics, Shahe Higher Education Park, Changping District, Beijing 102206, China.}
\address[label2]{The University of Sydney Business School, NSW 2006, Australia}
\address[label3]{School of Statistics and Mathematics, Central University of Finance and Economics, Beijing, China}


\begin{abstract}
  The practical importance of coherent forecasts in hierarchical forecasting has inspired many studies on forecast reconciliation. Under this approach, so-called base forecasts are produced for every series in the hierarchy and are subsequently adjusted to be coherent in a second reconciliation step. Reconciliation methods have been shown to improve forecast accuracy, but will, in general, adjust the base forecast of every series. However, in an operational context, it is sometimes necessary or beneficial to keep forecasts of some variables unchanged after forecast reconciliation. In this paper, we formulate reconciliation methodology that keeps forecasts of a pre-specified subset of variables unchanged or ``immutable''. In contrast to existing approaches, these immutable forecasts need not all come from the same level of a hierarchy, and our method can also be applied to grouped hierarchies. We prove that our approach preserves unbiasedness in base forecasts. Our method can also account for correlations between base forecasting errors and ensure non-negativity of forecasts.  We also perform empirical experiments, including an application to sales of a large scale online retailer, to assess the impacts of our proposed methodology.
\end{abstract}
\begin{keyword}
Forecasting\sep
Hierarchical time series \sep
Constrained optimization\sep
Unbiasedness\sep
Online retail
\end{keyword}

\end{frontmatter}

\newpage

\section{Introduction}

Hierarchical time series are characterized by aggregation, or more generally by linear constraints in the temporal and or cross-sectional domain.  They arise in many applications including the field of demand forecasting, where sales of individual products can be grouped into a hierarchy of sub categories and categories, or daily demand can be aggregated into weekly, monthly and annual demand. Forecasts of such hierarchies support supply chain management and production planning (\Citealp{syntetosSupplyChainForecasting2016}). Current state-of-the-art methods first generate forecasts of all variables in a hierarchy (known as ``base'' forecasts), and then ``reconcile'' these to ensure they respect aggregation constraints. These methods generally change the forecasts of all series in the hierarchy. However, in a number of operational settings, there may be practical reasons for reconciling forecasts while leaving forecasts of a subset of pre-specified series unchanged. This problem forms the subject of investigation of this paper.

Traditional approaches to forecasting hierarchical times series (bottom-up, top-down, middle-out) focused on producing forecasts of a single level of the hierarchy and then obtaining forecasts for other levels via aggregation, disaggregation or a combination of both. The pioneering work of \citet{hyndmanOptimalCombinationForecasts2011} proposed forecasting all variables in a hierarchy and then reconciling these variables via a least-square problem such that the new forecasts were coherent (i.e. aggregation constraints are satisfied). These methods have since been applied to temporal hierarchies (\Citealp{athanasopoulosForecastingTemporalHierarchies2017}), and extended to incorporate correlations between base forecasting errors (\Citealp{wickramasuriyaOptimalForecastReconciliation2019}). The theoretical properties of these methods have been explored through both a game theoretic \citep{vanervenGameTheoreticallyOptimalReconciliation2015} and geometric \citep{panagiotelisForecastReconciliationGeometric2021} lens, with the latter establishing that any method based on projections preserves the unbiasedness of base forecasts. Reconciliation has also been shown to achieve significant accuracy improvements in several real applications (\Citealp{yangReconcilingSolarForecasts2017, athanasopoulosForecastingTemporalHierarchies2017, zhangLeastSquaresbasedOptimal2018, jeonProbabilisticForecastReconciliation2019, bentaiebHierarchicalProbabilisticForecasting2020}).

It is important to note that in contrast to bottom up, top down or middle out methods, reconciliation will adjust the forecasts of all series in a hierarchy. Also while reconciliation improves forecast accuracy in the hierarchy overall, reconciled forecasts of individual series within the hierarchy may experience only negligible improvement, or even a deterioration in accuracy relative to base forecasts. \citet{nystrupDimensionalityReductionForecasting2021} discuss how this depends on error structure and the degree of incoherency while
\cite{athanasopoulosForecastingTemporalHierarchies2017} argue that the greatest gains in forecast accuracy usually occur for the most inaccurate base forecasts. Indeed, several case studies~\citep[e.g.][]{wickramasuriyaOptimalForecastReconciliation2019,hollymanUnderstandingForecastReconciliation2021a,pritulargaStochasticCoherencyForecast2021} show that the accuracy of the top level forecast worsen after the reconciliation step, which is of particular concern since practitioners often care most about the top level. To address this issue \citet{hollymanUnderstandingForecastReconciliation2021a} propose a reconciliation method that leaves the top-level forecast unchanged. Unlike traditional top-down methods based on disaggregation, their method is a projection, therefore preserves the unbiasedness of forecasts. They generalize their method so that forecasts at any level (and not just the top or bottom level) remain unchanged. Subsequent work by \citet{difonzoForecastCombinationBased2021} further explore the theoretical properties of the method proposed by \citet{hollymanUnderstandingForecastReconciliation2021a}.

We extend this literature by proposing a method for optimal reconciliation that keeps forecasts of a subset of \textit{series} unchanged or ``immutable''. In contrast to \citet{hollymanUnderstandingForecastReconciliation2021a} and \citet{difonzoForecastCombinationBased2021}, the immutable series in our proposed method may come from \textit{different levels} of the hierarchy. This may be particularly attractive when both the top-level forecasts and the forecasts of some key individual product lines are subject to judgmental adjustments by experts and therefore should not be altered by the reconciliation process. Another practical scenario where keeping forecasts from different levels immutable, is discussed in our application in Section~\ref{sec:application} where the amount of available training data is used to determine the set of immutable forecasts. The set of immutable forecasts cannot be chosen completely arbitrarily and conditions for selecting a valid set of immutable forecasts are also derived in this paper. Our method preserves the unbiasedness of forecasts, a result we prove in this paper.

Our proposed method also extends \citet{hollymanUnderstandingForecastReconciliation2021a} and \citet{difonzoForecastCombinationBased2021} in three more important ways. First, it is combined with the non-negative reconciliation approach in \citet{wickramasuriyaOptimalNonnegativeForecast2020} via the imposition of inequality constraints. This is particularly critical in our application to sales data which are non-negative. Second, the proposed method takes advantage of the full covariance matrix of forecasting errors, which improves forecasts by facilitating information sharing between nodes and levels (see, e.g.,\Citealp{wickramasuriyaOptimalForecastReconciliation2019, nystrupTemporalHierarchiesAutocorrelation2020}). Third, our approach is more general than the LCC method of \cite{hollymanUnderstandingForecastReconciliation2021a} in that it can be applied to grouped hierarchies, an example of which we provide in Section~\ref{sec:casestudy}.

The rest of the paper is organized as follows. In Section~\ref{sec:relatedwork}, we briefly review the relevant literature on forecast reconciliation. We then demonstrate our constrained forecast reconciliation method in Section~\ref{sec:method}. In Section~\ref{sec:simulations}, we conduct two Monte Carlo experiments to show the benefits of our methods under certain circumstances. In Section~\ref{sec:application}, we evaluate the performance of the proposed method with an application to the demand data in an online supermarket in China. Section~\ref{sec:casestudy}, demonstrates its use and benefits in a case study of the Wikipedia page views dataset, followed by concluding remarks in Section~\ref{sec:conclusions}. The code for reproducing the results is available at \url{https://github.com/AngelPone/chf}.

\section{Related work}
\label{sec:relatedwork}

This section briefly introduces the notation used in this paper and the methodological framework of forecast reconciliation.
For simplicity, we only provide a full exposition of the cross-sectional context hierarchical time series noting that the extension to temporal hierarchical time series is straightforward.

\subsection{Notation}

For a given hierarchy, let $n$ denote the total number of time series in the hierarchy, and $\boldsymbol{y}_t \in \mathbb{R}^n$ denotes a vector of observations at time $t$.
To preserve the full generality, we adopt the concepts of hierarchical time series and coherence proposed by \cite{panagiotelisForecastReconciliationGeometric2021}.
Let $\boldsymbol{b}_t\in \mathbb{R}^m$ denotes a vector of \textit{basis} time series.
In contrast with most of the literature for which  $\boldsymbol{b}_t$ denotes ``\textit{bottom}'' level series, in our case, $\boldsymbol{b}_t$ can consist of time series from multiple levels in the hierarchy.
$\boldsymbol{y}_t$ can be determined by right multiplying matrix $\boldsymbol{S}$ (a constant matrix of order $n \times m$) by $\boldsymbol{b}_t$. Here, $\bm{S}$ is not necessarily a \textit{summing matrix}, but its columns still span the subspace where forecasts are coherent. Thus, for observations at time $t$, we have
\[
\boldsymbol y_t = \boldsymbol{Sb}_t.
\]

To make this clearer consider a 3-variable hierarchy where $X=Y+Z$. Defining with basis series $[Y,~Z]'$ and $[X,~Y]'$ allow the hierarchy to be defines as

\[
\begin{pmatrix}X\\Y\\Z\end{pmatrix}=\begin{pmatrix}1&1\\1&0\\0&1\end{pmatrix}\begin{pmatrix}Y\\Z\end{pmatrix}\,\textrm{or}\,\begin{pmatrix}X\\Y\\Z\end{pmatrix}=\begin{pmatrix}1&0\\0&1\\1&-1\end{pmatrix}\begin{pmatrix}X\\Y\end{pmatrix}\,.
\]

Despite the fact that the definition on the left is almost always used, the existing reconciliation methods described in this section are invariant to the way the hierarchy is characterized. The reason why we consider characterizations such as that on the right will become clearer in Section~\ref{sec:method}.

We use the term \textit{determined} time series for the $n-m$ elements  in $\boldsymbol{y}_t$ that are not basis time series. A vector of $h$-step-ahead incoherent base forecasts  given observations up to time $T$
is denoted $\hat {\boldsymbol y}_T(h)$.
Let $\psi$ be a mapping that reconciles the base forecasts to be coherent, i.e., $\tilde {\boldsymbol y}_T(h) = \psi(\hat{\boldsymbol{y}}_T(h))$, where $\tilde {\boldsymbol y}_T(h)$ denotes the reconciled forecasts subjective to the aggregation constraints.
In this paper, we focus on the linear forecast reconciliation method, in which
\[
  \label{eq_SG}
\tilde {\boldsymbol y}_T(h) = \boldsymbol{SG}\hat {\boldsymbol y}_T(h).
\]
$\boldsymbol G$ is an $m \times n$ weighting matrix that combines the base forecasts to obtain reconciled forecasts for basis time series, i.e., $\tilde{\boldmath{b}}_T(h)$.
They are then mapped to the coherent forecasts of all levels by pre-multiplying the summing matrix $\boldsymbol{S}$.
The choice of $\boldsymbol {G}$ depends on the estimator used in forecast reconciliation.

\subsection{Forecast reconciliation}

Although motivated in different ways, most forecast reconciliation methods find the reconciliation weights $\bm{G}$ by solving the following optimization problem:

\begin{equation}
\begin{aligned}
\underset{\bm{G}}{\min}\:(\hat{\bm{y}}_T(h)-\bm{S}\bm{G}\hat{\bm{y}}_T(h))'\bm{W}(\hat{\bm y}_T(h)-\bm{S}\bm{G}\hat{\bm{y}}_T(h))\\
\textrm{s.t.}\:\:\bm{G}\bm{S}=\bm{I},\\
\label{eq:stdreco}
\end{aligned}
\end{equation}
which for a given $\bm{W}$ is solved by $\hat{\bm{G}}=(\boldsymbol{S}^\prime\boldsymbol{W}\boldsymbol{S})^{-1}\boldsymbol{S}' \boldsymbol{W}$. The objective function in Equation~\ref{eq:stdreco} can be more compactly written as $(\hat{\bm{y}}-\tilde{\bm{y}})'\bm{W}(\hat{\bm{y}}-\tilde{\bm{y}})$, a convention we follow for the remainder of the paper. Different reconciliation methods correspond to different choices for $\bm{W}$. For example, the original \cite{hyndmanOptimalCombinationForecasts2011} paper uses $\bm{I}$, \cite{vanervenGameTheoreticallyOptimalReconciliation2015} and \cite{athanasopoulosForecastingTemporalHierarchies2017} assume a diagonal $\bm{W}$, while \cite{wickramasuriyaOptimalForecastReconciliation2019} prove that setting $\bm{W}$ to the inverse covariance matrix of forecast errors has a number of attractive theoretical properties. Alternative estimation strategies for the covariance matrix have been proposed by \citet{nystrupTemporalHierarchiesAutocorrelation2020,nystrupDimensionalityReductionForecasting2021} and \cite{pritulargaStochasticCoherencyForecast2021}. An alternative approach is to directly regularize the elements of $\bm{G}$ as proposed by \cite{taiebSparseSmoothAdjustments2017} and \cite{bentaiebRegularizedRegressionHierarchical2019a}, but unlike the general optimization problem given in~\ref{eq:stdreco}, these approaches do depend on the way the basis series and $\bm{S}$ matrix are defined.

An important modification to Equation~\ref{eq:stdreco} is given by
\cite{wickramasuriyaOptimalNonnegativeForecast2020} who restrict the reconciled forecasts to be non-negative, a particularly important feature in applications such as demand forecasting. They implemented the proposed method by adding non-negativity constraints to the original least-square minimization problem and solved this quadratic programming problem using three standard QP algorithms.

\begin{equation}
\begin{aligned}
\underset{\bm{G}}{\min}\:(\hat{\bm{y}}-\tilde{\bm{y}})'\bm{W}(\hat{\bm y}-\tilde{\bm{y}})\\
\textrm{s.t.}\:\:\tilde{\bm{b}}\geq 0.
\label{eq:stdnnreco}
\end{aligned}
\end{equation}\,

While the algorithms they used speed up the computation, dealing with high-dimensional hierarchies can be time-consuming. In section~\ref{sec:method} we extend this approach to the case with immutable forecasts, which can improve computational speed by keeping immutable forecasts of nodes with forecast error variance and thus reducing the effective dimension of the hierarchy.

\subsection{Forecast reconciliation with immutable forecasts}

\citet{hollymanUnderstandingForecastReconciliation2021a} proposed the level-conditional coherent (LCC) forecasts, obtained by minimizing the variance of the reconciled bottom level forecasts, while in the meantime, ensuring the summation of reconciled bottom-level forecasts equal the base forecasts of some specified level.


Our method generalizes the LCC method in three ways. First, our proposed method is able to keep multiple nodes from different levels immutable during reconciliation, while for LCC all immutable series must be from the same level. Our application in Section~\ref{sec:application} provides one such scenario where nodes from different levels must be kept immutable. Secondly, our method allows for non-diagonal $\bm{W}$, allowing information in the forecast error covariance matrix to be accounted for, thus enhancing the performance of reconciled forecasts.
Third, our approach generalizes to grouped hierarchies.
More importantly, from the mathematical point of view, the LCC approach reconciles the base forecasts within a low-level hierarchy - the conditional level and the bottom level - using the variance scaling approach (\Citealp{hyndmanFastComputationReconciled2016a,wickramasuriyaOptimalForecastReconciliation2019}) and keeping the conditional level immutable, which is also pointed out by \citet{difonzoForecastCombinationBased2021}. Consequently, LCC approach would only incorporate information of bottom level. However, information of other levels is also important. Our proposed method would consider the covariance information of all the mutable nodes in the hierarchy.

\begin{figure}[t]
  \centering
  \resizebox{0.5\textwidth}{!}{\includegraphics{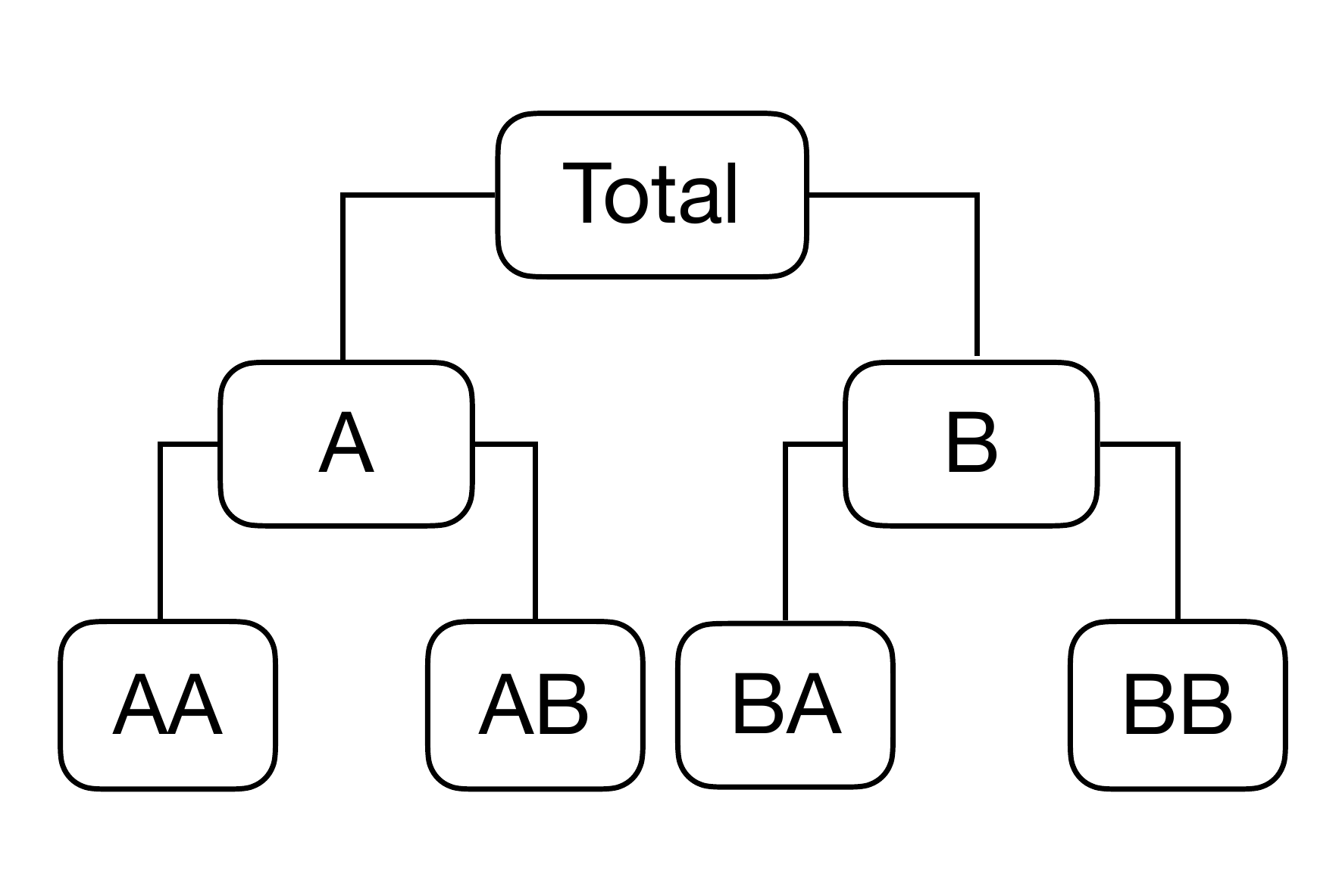}}
  \caption{\label{hierarchy} A simple three-level hierarchy}
\end{figure}

To emphasize the difference between LCC and our proposed method, we consider a three-level hierarchy shown in Figure~\ref{hierarchy}.
To keep $\hat y_{Total}$ immutable, our proposed method tries to optimize the following problem:
\begin{equation}
  \begin{aligned}
  \min_{\tilde{\bm{y}}}(\hat{\bm{y}} - \tilde{\bm{y}})'\bm{W}^{-1}(\hat{\bm{y}} - \tilde{\bm{y}}) \\
  \textrm{s.t.} \:\: \tilde{y}_{Total} = \hat y_{Total},
  \end{aligned}
\end{equation}
where $\bm{W}$ can be any estimators of the covariance matrix of base forecast errors. Thus, our proposed method can potentially incorporate information from all mutable levels. However, the LCC method uses a diagonal estimator of the covariance matrix, $\textrm{diag}([0, 0, 0, \sigma_{AA}^2,\sigma_{AB}^2,\allowbreak \sigma_{BA}^2,\sigma_{BB}^2])$.
The estimator entirely ignores information of middle level and covariance.
Furthermore, if $y_{A}$ is set to be immutable, LCC would only consider the small hierarchy $\bm{y} = [\bm{y}_{A}, \bm{y}_{AA}, \bm{y}_{AB}]'$ (\Citealp{difonzoForecastCombinationBased2021}) and ignore information from other nodes, which can be captured by our proposed method.
Besides, LCC can not be used when a subset of bottom-level series is immutable.


\section{Forecast reconciliation with equality constraints}
\label{sec:method}

We now introduce our method for keeping forecasts of a chosen set of variables, not all necessarily from the same level, unchanged after reconciliation, i.e. immutable. We denote the immutable series with the $k$-vector $\bm{u}_t$. The basis time series must be chosen so that the immutable series are a sub-vector of the basis series. Without loss of generality, we assume that these are the last $k$ elements of $\bm{b}_t$. The first $m - k$ elements of $\bm{b}_t$ are mutable time series, which we denote $\bm{v}_t$. The determined time series (which by construction are also mutable) are denoted $\bm{w}_t$. The hierarchy of base forecasts is thus stacked as, $\hat {\boldsymbol{y}}_t = (\hat {\boldsymbol{w}}_t, \hat {\boldsymbol{v}}'_t, \hat {\boldsymbol{u}}'_t)'$. A summary of the terminology and notation is provided in Table~\ref{tab:defs} where subscripts are dropped for ease of exposition.

\begin{table}[!h]
{\scriptsize
	\begin{centering}
		\caption{Summary of terminology and notation used for different series}
	\begin{tabular}{rll}
    \toprule
		Term &Notation &Description\\
    \midrule
		Full hierarchy	&$\bm{y}=\bm{(w',v',u')'}$& All series.\\
		Immutable & $\bm{u}$& Series with forecasts that do not change after reconciliation.\\
		Mutable & $\bm{(w',v')'}$&Series with forecasts that do change after reconciliation.\\
		Basis & $\bm{b}=\bm{(v',u')'}$& Series used to define hierarchy, not necessarily the same as bottom level.\\
		Determined & $(\bm{u'},\bm{w'})'$ & Series that can be determined if values of basis level and constraints are known.\\
    \bottomrule
	\end{tabular}
\label{tab:defs}
\end{centering}
}
\end{table}

 The summing matrix can be partitioned as
\[
  \boldsymbol{S} = \begin{bmatrix*}[l]
    \boldsymbol{S}_1 & \boldsymbol{S}_2 \\
    \boldsymbol{I}_{m-k} & \boldsymbol{0}_{(m-k)\times k} \\
    \boldsymbol{0}_{k \times (m-k)} & \boldsymbol{I}_{k\times k}
  \end{bmatrix*},
\]

where $\boldsymbol{S}_1$ and $\boldsymbol{S}_2$ are $(n-m)\times (m-k)$ and $(n-m)\times (k)$ matrices respectively, $\boldsymbol{0}_{a\times b}$  is an $a\times b$ matrix of zeros and $\boldsymbol{I}_a$ is an $a\times a$ identity matrix. Immutable forecasts are found as the solution to the following optimization problem can be written as
\begin{equation}
  \begin{aligned}
  \min\limits_{\tilde{\boldsymbol{y}}} (\hat{\boldsymbol{y}} - \tilde{\boldsymbol{y}})^\prime \boldsymbol{W}^{-1}& (\hat{\boldsymbol{y}}-\tilde{\boldsymbol{y}}) \\
  \textrm{s.t.} \quad \tilde{\boldsymbol{u}} &= \hat{\boldsymbol{u}}, \\
   \boldsymbol{S_1} \tilde{\boldsymbol{v}} + \boldsymbol{S}_2 \tilde{\boldsymbol{u}} &= \tilde{\boldsymbol{w}},
  \end{aligned}
\end{equation}
where $\boldsymbol{W}$ is the covariance matrix of the base forecast errors.
The first constraint guarantees that immutability of $\bm{u}$.
The second constraint is the aggregation constraint.
By substituting $\tilde{\boldsymbol{w}}$ into the objective function, the original constrained optimization problem becomes the following unconstrained problem:
\begin{equation}
  \label{eq2}
  \min\limits_{\tilde{\boldsymbol{v}}} (\check{\boldsymbol{\nu}} - \check{\boldsymbol{S}_1} \tilde{\boldsymbol{v}})^\prime \boldsymbol{W}_{\boldsymbol{\nu}}^{-1}(\check{\boldsymbol{\nu}} - \check{\boldsymbol{S}_1} \tilde{\boldsymbol{v}}),
\end{equation}
where \[\check{\boldsymbol{S}_1} = \left[\begin{matrix}
  \boldsymbol{S}_1 \\
  \boldsymbol{I}_{m-k}
\end{matrix}\right], \check{\boldsymbol{\nu}} = \left[\begin{matrix}
  \hat{\boldsymbol{w}} - \boldsymbol{S}_2 \hat{\boldsymbol{u}} \\
  \hat{\boldsymbol{v}}
\end{matrix}\right],\]
and $\boldsymbol{W}_{\boldsymbol{\nu}}$ is the covariance matrix of base forecast errors of the mutable series. This a generalized least squares problem, similar to that in Equation~\ref{eq:stdreco} with solution
\begin{equation}
  \label{eq3}
  \tilde{\boldsymbol{v}} = (\check{\boldsymbol{S}}_1^\prime\boldsymbol{W}_{\boldsymbol{\nu}}^{-1}\check{\boldsymbol{S}}_1)^{-1}\check{\boldsymbol{S}}_1^\prime\boldsymbol{W}_{\boldsymbol{\nu}}^{-1}\check{\boldsymbol{\nu}}.
\end{equation}
Denote $\hat{\boldsymbol{G}}_1 = (\check{\boldsymbol{S}}_1^\prime\boldsymbol{W}_{\boldsymbol{\nu}}^{-1}\check{\boldsymbol{S}}_1)^{-1}\check{\boldsymbol{S}}_1^\prime\boldsymbol{W}_{\boldsymbol{\nu}}^{-1}$. Then we have
\[
  \tilde{\boldsymbol{b}} = \left[\begin{matrix}\tilde{\boldsymbol{v}} \\ \hat{\boldsymbol{u}}\end{matrix}\right]=\left[\begin{matrix*}[l]
    \hat{\boldsymbol{G}}_1 & \boldsymbol{0}_{(m-k)\times k} \\
    \boldsymbol{0}_{k \times (n-k)} & \boldsymbol{I}_{k}
  \end{matrix*}\right](\boldsymbol{I}_n - \left[\begin{matrix*}[l]
    \boldsymbol{0}_{(n-m)\times (n-k)} & \boldsymbol{S}_2 \\
    \boldsymbol{0}_{m\times (n-k)} & \boldsymbol{0}_{m\times k}
  \end{matrix*}\right])\left[\begin{matrix}
    \hat{\boldsymbol{w}} \\
    \hat{\boldsymbol{v}} \\
    \hat{\boldsymbol{u}}
  \end{matrix}\right]   = \hat{\boldsymbol{G}}\hat{\boldsymbol{y}}.
  \label{eq:optgim}
\]
The reconciled forecasts of the hierarchy can be obtained through $\tilde{\boldsymbol{y}} = \boldsymbol{S}\tilde{\boldsymbol{b}} = \boldsymbol{S}\hat{\boldsymbol{G}}\hat{\boldsymbol{y}}$.

\begin{thm}
	If the vector of base forecasts $\hat{\bm{y}}$ is unbiased, then the reconciled forecasts $\boldsymbol{S}\hat{\boldsymbol{G}}\hat{\boldsymbol{y}}$ with $\hat{\boldsymbol{G}}$ given by Equation~\ref{eq:optgim} are also unbiased.
\end{thm}

\begin{proof}
	See \hyperref[appendixB]{Appendix}.
\end{proof}

The covariance matrix of base forecast errors $\boldsymbol{W}$ is unidentifiable and can be estimated in terms of historical base forecast errors.
In this paper, we use four estimators also used in \cite{wickramasuriyaOptimalForecastReconciliation2019}, which are OLS estimator, WLS applying variance scaling, WLS applying structural scaling and shrinkage estimator.
They are denoted as $\text{OLS}$, $\text{WLS}_{v}$, $\text{WLS}_{s}$ and $\text{MinT}(\text{Shrinkage})$, respectively.

Our proposed method for immutable forecasts can be combined with non-negative reconciliation proposed by \cite{wickramasuriyaOptimalNonnegativeForecast2020} by imposing nonnegativity constraints into Equation~\eqref{eq2}.
The optimization problem becomes
\begin{equation}
  \begin{aligned}
  \min\limits_{\tilde{\boldsymbol{v}}} (\check{\boldsymbol{\nu}} - \check{\boldsymbol{S}_1} \tilde{\boldsymbol{v}})^\prime \boldsymbol{W}_{\boldsymbol{\nu}}^{-1}(\check{\boldsymbol{\nu}} - \check{\boldsymbol{S}_1} \tilde{\boldsymbol{v}}) \\
  \textrm{s.t.}  \quad \check{\boldsymbol{v}} \geq 0.
  \end{aligned}
  \label{eq:opt_nn}
\end{equation}
This quadratic programming problem has a unique global solution and can be solved by \pkg{quadprog} package for \proglang{R} when $\tilde{\boldsymbol{v}}$ has low dimension.
When $\tilde{\boldsymbol{v}}$ is of high dimension, \cite{wickramasuriyaOptimalNonnegativeForecast2020} suggested using the block principal pivoting (BPV) algorithm, projected conjugate gradient (PCG) algorithm, or scaled gradient projection algorithm to accelerate the solution.

Note that the solution to Equation~\eqref{eq:opt_nn} will not still be unbiased when imposing nonnegativity constraints. Also, care must be taken in using this algorithm, since although the algorithm guarantees non-negativity of mutable forecasts, negative base forecasts of immutable series will not be changed after reconciliation. In these cases, reconciled forecasts of some determined series may also be negative. Therefore, in applications where non-negative forecasts are required, we recommend that steps should be taken to ensure base forecasts are not negative.

\subsection{Conditions for selecting a set of immutable forecasts}
\label{sec:immutcond}

Any subset of series in the basis set can be immutable. However, the basis set can not be chosen arbitrarily for a given hierarchy. Let $\bm{y} = \bm{Sb}$ be any valid representation of the hierarchy (e.g., $\bm{b}$ can be bottom-level series). Let $\bm{b}^* = (y_{j_1},\dots,y_{j_m})'$ be a candidate set of basis series, where $j_1,\dots,j_m$ are the indices of the elements of $\bm{y}$. Let $\bm{S}_{\{j\}}$ be a square matrix formed by taking the rows of $\bm{S}$ corresponding to the indices $j_1,\dots, j_m$.

\begin{thm}
  A candidate set of basis series $\bm{b}^*$ is valid if $\bm{S}_{\{j\}}$ is invertible.
\end{thm}

\begin{proof}
	See \hyperref[appendixB]{Appendix}.
\end{proof}

\section{Monte carlo simulations}
\label{sec:simulations}

In order to demonstrate the applicability of the method proposed in Section ~\ref{sec:method}, we carry out two Monte Carlo experiments. In Section~\ref{study1}, we focus on the situation where we know the underlying data generating process (DGP) of the top level series, but have limited information about the disaggregated series. In Section~\ref{study2}, we concentrate on the scenario, where the disaggregated series are noisy while the aggregated series are less noisy compared to the disaggregated series due to the smoothing effect of aggregation. We consider a simple hierarchy as shown in Figure~\ref{hierarchy}. There are four time series at the bottom level and each two of them are aggregated to obtain the middle level series. Summing the middle level gives the top level.

\subsection{Scenario I: model misspecification in disaggregated levels}
\label{study1}

\subsubsection{Simulation setup}

Firstly, we want to simulate a scenario where we have limited information about disaggregated levels, so that the model is misspecified, while the top level is easier to model due to aggregation. We simulate the series in a bottom-up manner. The bottom-level time series are generated using the basic structural time series model
\[\boldsymbol{b_t} = \boldsymbol{\mu_t}+\boldsymbol{s}_t + \boldsymbol{\eta}_t,
\]
where $\boldsymbol{\mu}_t$, $\boldsymbol{s}_t$ and $\boldsymbol{\eta}_t$ are trend, seasonal and error terms, respectively. They are simulated using
\[\begin{array}{ll}\boldsymbol{\mu}_t = \boldsymbol{\mu}_{t-1}+ \boldsymbol v_t + \boldsymbol e_t, &\boldsymbol e_t \sim \mathcal{N}(\boldsymbol 0, \sigma^2_{e}\boldsymbol I_4), \\

\boldsymbol v_t = \boldsymbol v_{t-1} + \boldsymbol  \varepsilon_t, &\boldsymbol \varepsilon_t \sim \mathcal{N}(\boldsymbol 0, \sigma^2_{\varepsilon}\boldsymbol I_4), \\

\boldsymbol s_t = -\sum_{i=1}^{l-1} \boldsymbol s_{t-i} + \boldsymbol \omega_t, &\boldsymbol \omega_t \sim \mathcal{N}(0,\sigma^2_\omega \boldsymbol I_4),

\end{array}
\]
where $\boldsymbol e_t$, $\boldsymbol\varepsilon_t$ and $\boldsymbol \omega_t$ are independent of each other. $l$ is the length of seasonality period. We set $l=12$ for monthly data. We follow the parameter settings in Section 3.4 of \cite{wickramasuriyaOptimalForecastReconciliation2019}, and set  $\sigma_e^2 = 2, \sigma^2_\varepsilon=0.007$ and $\sigma_\omega^2 = 7$. The initial values for $\boldsymbol \mu_0, \boldsymbol v_0, \boldsymbol s_0,\boldsymbol s_1, \dots, \boldsymbol s_{11}$ are independently generated from a multivariate normal distribution with mean zero and covariance matrix $\boldsymbol \Sigma_0 = \boldsymbol I_4$. The noise terms $\boldsymbol\eta_t$ are generated from ARIMA($p$, $0$, $q$) with $p$ and $q$ taking value of 0 and 1 with equal probability. The coefficients of AR and MA component in the ARIMA process are sampled from $[0.5, 0.7]$ uniformly. To simulate the dependence structure between time series in the bottom level, the ARIMA DGPs have a contemporaneous error term with covariance matrix
\[\boldsymbol\Sigma_1= \left[\begin{array}{cccc} 3 & -2 & 0 & 0 \\

-2 & 3 & 0 & 0 \\

0 & 0 &3 & -1 \\

0 & 0 & -1 & 3

\end{array}\right].\]
We generate $T=324$ observations for each time series, and the final $h=24$ points are used as test data for evaluation. Base forecasts are produced in two different ways. In the first way, we use Exponential Smoothing (ETS) models for all levels. In the second way, we use ETS models for the top levels while Autoregressive Integrated Moving Average (ARIMA) models are used for the remaining levels. The ARIMA and ETS models are implemented using the default parameters of \code{auto.arima()} and \code{ets()} functions in the \pkg{forecast} package for \proglang{R} \citep{hyndmanAutomaticTimeSeries2008}.

\subsubsection{Forecasting results}

Using the parameters described above, the simulation process is repeated 1000 times. The average root mean squared error (RMSE) of each aggregation level and their mean values are reported. The upper panel of Table~\ref{study1_res} shows the results using the ETS base forecasts, and the lower panel shows the results using ETS base forecasts for the top level and ARIMA base forecasts for the other levels.

\begin{table}[]
  \caption{\label{study1_res} Out of sample forecasting accuracies for the simulated data in scenario I (model misspecification in disaggregated levels). The upper panel (``ETS'') shows the results using the ETS base forecasts, and the lower panel (``ETS + ARIMA'') shows the results using ETS base forecasts for the top level and ARIMA base forecasts for the other levels. Subcolumns ``U'' and ``C'' show forecast accuracies of unconstrained and constrained forecast reconciliation using different covariance matrices - OLS, $\text{WLS}_s$, $\text{WLS}_v$ and  MinT (Shrinkage). The accuracy of base forecasts is also shown for comparison.}
  \resizebox{\textwidth}{!}{
  \begin{tabular}{lrrrrrrrrr}
    \toprule
    & \multicolumn{9}{c}{ETS} \\ \cmidrule(lr){2-10}
        \multirow{2}{*}{Level} & \multirow{2}{*}{Base} & \multicolumn{2}{c}{OLS}                                             & \multicolumn{2}{c}{$\text{WLS}_s$}                                          & \multicolumn{2}{c}{$\text{WLS}_v$}                                          & \multicolumn{2}{c}{MinT (Shrinkage)}                                 \\
       & \multicolumn{1}{c}{}     & \multicolumn{1}{c}{U} & \multicolumn{1}{c}{C} & \multicolumn{1}{c}{U} & \multicolumn{1}{c}{C} & \multicolumn{1}{c}{U} & \multicolumn{1}{c}{C} & \multicolumn{1}{c}{U} & \multicolumn{1}{c}{C} \\ \cmidrule(lr){3-4} \cmidrule(lr){5-6} \cmidrule(lr){7-8} \cmidrule(lr){9-10}

0 &        21.4003 & 21.3741 & 21.4003 & 21.3893 & 21.4003 & 21.4204 & 21.4003 & 21.4006 & 21.4003\\
1 & 11.9528 & 11.9269 & 11.9393 & 11.9255 & 11.9316 & 11.9347 & 11.9270 & 11.9165 & 11.9176\\
2 & 7.1655 & 7.1424 & 7.1474 & 7.1418 & 7.1440 & 7.1454 & 7.1417 & 7.1398 & 7.1392\\
Average & 13.5062 & 13.4811 & 13.4956 & 13.4855 & 13.4920 & 13.5002 & 13.4897 & 13.4856 & 13.4857\\\midrule

        & \multicolumn{1}{c}{}     & \multicolumn{1}{c}{}              & \multicolumn{1}{c}{}            & \multicolumn{1}{c}{}              & \multicolumn{1}{c}{}            & \multicolumn{1}{c}{}              & \multicolumn{1}{c}{}            & \multicolumn{1}{c}{}              & \multicolumn{1}{c}{} \\

      & \multicolumn{9}{c}{ETS + ARIMA} \\ \cmidrule(lr){2-10}
      \multirow{2}{*}{Level}  & \multirow{2}{*}{Base} & \multicolumn{2}{c}{OLS}                                             & \multicolumn{2}{c}{$\text{WLS}_s$}                                          & \multicolumn{2}{c}{$\text{WLS}_v$}                                          & \multicolumn{2}{c}{MinT (Shrinkage)}                                 \\
      & \multicolumn{1}{c}{}     & \multicolumn{1}{c}{U} & \multicolumn{1}{c}{C} & \multicolumn{1}{c}{U} & \multicolumn{1}{c}{C} & \multicolumn{1}{c}{U} & \multicolumn{1}{c}{C} & \multicolumn{1}{c}{U} & \multicolumn{1}{c}{C} \\ \cmidrule(lr){3-4} \cmidrule(lr){5-6} \cmidrule(lr){7-8} \cmidrule(lr){9-10}
0&        21.4003 & 21.4422 & 21.4003 & 21.6658 & 21.4003 & 21.9257 & 21.4003 & 22.1046 & 21.4003\\
1&       12.6553 & 12.1994 & 12.1921 & 12.2582 & 12.1625 & 12.3544 & 12.1495 & 12.4509 & 12.1726\\
2&        7.7049 & 7.5549 & 7.5562 & 7.5745 & 7.5440 & 7.6079 & 7.5359 & 7.6255 & 7.5303\\
Average&        13.9202 & 13.7322 & 13.7162 & 13.8328 & 13.7022 & 13.9627 & 13.6952 & 14.0603 & 13.7010\\\bottomrule

        & \multicolumn{1}{l}{}     & \multicolumn{1}{l}{}              & \multicolumn{1}{l}{}            & \multicolumn{1}{l}{}              & \multicolumn{1}{l}{}            & \multicolumn{1}{l}{}              & \multicolumn{1}{l}{}            & \multicolumn{1}{l}{}              & \multicolumn{1}{l}{}
      \end{tabular}}
  \end{table}

Table~\ref{study1_res} shows the out of sample forecasting results for the simulated data using four different covariance matrix estimators.

We use the following taxonomy for different methods. We use ``C'' to denote the method described in section~\ref{sec:method} with immutability constraints (top level immutable), while ``U'' denotes reconciliation without immutability constraints as described in Section~\ref{sec:relatedwork}. We also consider three weighting schemes. First, OLS sets $\bm{W}=\bm{I}$. Second, $\textrm{WLS}_s$ refers to the structural scaling of \cite{athanasopoulosForecastingTemporalHierarchies2017}, where $\bm{W}$ is a diagonal matrix with weights is inversely proportional to the number of series used to form an aggregate. Third, $\textrm{WLS}_v$ refers to the case were $\bm{W}$ is a diagonal matrix with weights is inversely proportional to the forecast error variance of each series.  A full estimate of the covariance matrix is not pursued here since the number of training observations is different for each time series.

We can see that, even though the regular forecast reconciliation improves the forecasting accuracy of disaggregated series, it may decrease the forecasting accuracy of the top level, especially when the disaggregated series forecasts are generated by ARIMA models. Keeping the base forecasts of the top level unchanged maintains the accuracy in the top level and shows extra improvements in the disaggregated levels. This simulation study corresponds to a common scenario in practice, in which the disaggregated series are hard to model while the aggregated series is easier to fit.

\subsection{Scenario II: exploring the smoothing effect of aggregation}
\label{study2}

\subsubsection{Simulation setup}
Another common scenario in practice is that the lower level series are highly volatile and have a low signal-to-noise ratio, while these fluctuations are smoothed out due to aggregation in the upper levels. We implement the simulation based on the work of \cite{vanervenGameTheoreticallyOptimalReconciliation2015} for the hierarchy shown in Figure~\ref{hierarchy}. The bottom level series are generated as described in Section~\ref{study1}. A noise component is then added to make the bottom level series noisier than the aggregated series. Following \cite{wickramasuriyaOptimalForecastReconciliation2019}, the bottom-level series are generated using
\[
  \begin{aligned}
    y_{AA, t} &= z_{AA, t} - v_t - 0.5 \omega_t, \\
    y_{AB, t} &= z_{AB, t} + v_t - 0.5\omega_t,\\
    y_{BA, t} &= z_{BA, t} - v_t + 0.5\omega_t, \\
    y_{BB, t} &= z_{BB, t} + v_t - 0.5\omega_t,
  \end{aligned}
\]
where $z_{AA,t}, z_{AB,t}, z_{BA,t}, z_{BB,t}$ are bottom-level series generated as described in Section~\ref{study1}, $v_t\sim \mathcal{N}(0, 10)$ and $\omega_t \sim \mathcal{N}(0, 9)$ are independent white noise processes. The bottom-level series are then added up to obtain aggregated series.

\subsubsection{Forecasting results}

Using the same parameters as Section~\ref{study1}, we repeat the simulation process 1000 times. As shown in Table~\ref{study2_res}, in this simulation study where the lower level series are noisier than Scenario I,  the unconstrained reconciliation methods always perform worse than the base forecasts for the top level, irrespective of whether ETS or ``ETS + ARIMA'' is used. Keeping the top level unchanged shows promising accuracy improvements compared to base forecasts and the unconstrained forecast reconciliation methods. Note that keeping the top-level base forecasts immutable in our experiments can be considered a top-down method. However, any level or multiple nodes from different hierarchical levels can be constrained in practice, which shows high flexibility of the proposed method.

\begin{table}[]
  \caption{\label{study2_res} Out of sample forecasting accuracies for the simulated data in scenario II (exploring the smoothing effect of aggregation). The upper panel (``ETS'') shows the results using the ETS base forecasts, and the lower panel (``ETS + ARIMA'') shows the results using ETS base forecasts for the top level and ARIMA base forecasts for the other levels. Subcolumns ``U'' and ``C'' show forecast accuracies of unconstrained and constrained forecast reconciliation using different covariance matrices - OLS, $\text{WLS}_s$, $\text{WLS}_v$ and  MinT (Shrinkage). The accuracy of base forecasts is also shown for comparison.}
  \resizebox{\textwidth}{!}{
  \begin{tabular}{lrrrrrrrrr}
    \toprule
    & \multicolumn{9}{c}{ETS} \\ \cmidrule(lr){2-10}
        \multirow{2}{*}{Level} & \multirow{2}{*}{Base} & \multicolumn{2}{c}{OLS}                                             & \multicolumn{2}{c}{$\text{WLS}_s$}                                          & \multicolumn{2}{c}{$\text{WLS}_v$}                                          & \multicolumn{2}{c}{MinT (Shrinkage)}                                 \\
       & \multicolumn{1}{c}{}     & \multicolumn{1}{c}{U} & \multicolumn{1}{c}{C} & \multicolumn{1}{c}{U} & \multicolumn{1}{c}{C} & \multicolumn{1}{c}{U} & \multicolumn{1}{c}{C} & \multicolumn{1}{c}{U} & \multicolumn{1}{c}{C} \\ \cmidrule(lr){3-4} \cmidrule(lr){5-6} \cmidrule(lr){7-8} \cmidrule(lr){9-10}
0&      21.209 & 21.233 & 21.209 & 21.348 & 21.209 & 21.405 & 21.209 & 21.271 & 21.209\\
1&      12.611 & 12.526 & 12.513 & 12.549 & 12.486 & 12.571 & 12.483 & 12.459 & 12.429\\
2&      8.487 & 8.382 & 8.377 & 8.391 & 8.366 & 8.400 & 8.366 & 8.357 & 8.346\\
Average&      14.102 & 14.047 & 14.033 & 14.096 & 14.020 & 14.126 & 14.019 & 14.029 & 13.995\\\midrule

        & \multicolumn{1}{l}{}     & \multicolumn{1}{l}{}              & \multicolumn{1}{l}{}            & \multicolumn{1}{l}{}              & \multicolumn{1}{l}{}            & \multicolumn{1}{l}{}              & \multicolumn{1}{l}{}            & \multicolumn{1}{l}{}              & \multicolumn{1}{l}{} \\

        & \multicolumn{9}{c}{ETS + ARIMA} \\ \cmidrule(lr){2-10}
        \multirow{2}{*}{Level} & \multirow{2}{*}{Base} & \multicolumn{2}{c}{OLS}                                             & \multicolumn{2}{c}{$\text{WLS}_s$}                                          & \multicolumn{2}{c}{$\text{WLS}_v$}                                          & \multicolumn{2}{c}{MinT (Shrinkage)}                                 \\
           & \multicolumn{1}{c}{}     & \multicolumn{1}{c}{U} & \multicolumn{1}{c}{C} & \multicolumn{1}{c}{U} & \multicolumn{1}{c}{C} & \multicolumn{1}{c}{U} & \multicolumn{1}{c}{C} & \multicolumn{1}{c}{U} & \multicolumn{1}{c}{C} \\ \cmidrule(lr){3-4} \cmidrule(lr){5-6} \cmidrule(lr){7-8} \cmidrule(lr){9-10}
0&21.2092 & 21.3563 & 21.2092 & 21.6417 & 21.2092 & 21.8059 & 21.2092 & 21.9663 & 21.2092\\
1&13.2206 & 12.7756 & 12.7188 & 12.8465 & 12.6751 & 12.9120 & 12.6724 & 12.9523 & 12.6451\\
2&8.8454 & 8.6479 & 8.6287 & 8.6732 & 8.6126 & 8.6979 & 8.6131 & 8.7076 & 8.5982\\
Average&14.4251 & 14.2599 & 14.1856 & 14.3871 & 14.1656 & 14.4720 & 14.1649 & 14.5421 & 14.1509\\\bottomrule
        & \multicolumn{1}{l}{}     & \multicolumn{1}{l}{}              & \multicolumn{1}{l}{}            & \multicolumn{1}{l}{}              & \multicolumn{1}{l}{}            & \multicolumn{1}{l}{}              & \multicolumn{1}{l}{}            & \multicolumn{1}{l}{}              & \multicolumn{1}{l}{}
      \end{tabular}}
  \end{table}

\section{Application: forecasting demand for promotion events in a major Chinese online retailer}
\label{sec:application}

Demand forecasting is crucial in supply chain management, especially for the e-commerce industry. Forecasts drive order, logistics and inventory management for large scale online retailers. Products and their sales naturally form cross-sectional hierarchies according to product categories. It is usually essential to forecast series at different levels, and reconcile the forecasts to make them coherent. In this section we discuss strategies for choosing immutable series is such hierarchies and apply these ideas to real sales data from a major Chinese online retailer.

First, we may choose to series for which demand is intermittent to have immutable forecasts. For large online retailers, such intermittent SKUs are common, and it may make sense to forecast demand for such SKUs as zero. Reconciliation however will typically change the values of intermittent forecasts. Applying our proposed method while keeping intermittent series immutable provides a simple but elegant solution to this problem. This also effectively reduces the dimension of hierarchy from thousands to hundreds, while at the same time achieving coherent forecasts.

A second strategy for choosing immutable series may be motivated by promotions. Retail sales have been shown to be significantly affected by promotion events, such as Black Friday in North America and the ``11.11'' online shopping festival in China. Using promotion information as exogenous variables is essential to improve  forecasts during such promotional events. However, a common issue in e-commerce platforms is that new products emerge frequently meaning that the training sample available for different SKUs will be of varying lengths. In these cases, forecasting sales of time series of newer products with a smaller number of observations using univariate forecasting methods can be challenging since they have insufficient observations, especially during promotional periods. One way to solve this issue is to use a cross-learning technique that trains a global forecasting model using all time series in the dataset and transfer knowledge about the promotion effect from other time series. When more training observations are used for older products compared to newer products we can expect the trained global models to produce better forecasts for older products. This motivates keeping the forecasts of older products immutable while still using reconciliation to cross-learn the promotion effect for newer products.

\subsection{Dataset}

The sales data are obtained from the category ``food'' in the Chinese online retailer (who cannot be identified since the data are proprietary), which consists of 40 subcategories and 1905 items. The data were daily collected from 2019-01-01 to 2021-09-12, but not every time series starts from 2019-01-01 due to the emergence of new products. Figure~\ref{total_sale} shows the total sales of the ``food'' category. There are several spikes at specific dates in each year, which are caused by large-scale promotional events that would affect the sales of most products, such as the famous ``11.11'' and ``12.12'' shopping festivals in China. The dataset contains information on these events, including their names, start and end dates, and a variable measuring the strength of the promotion strength on each day. Promotion strength can be simply understood as related to the extent to which product prices are discounted, and is divided into multiple levels. Each promotional event is divided into multiple stages with different promotion strengths. For example, the ``11.11'' event in 2020 starts from  2020-10-21 to 2020-11-13. This time period is divided into five stages and the highest discount occurs on November 11th, which has the highest total sales in 2020. The dataset also contains a calendar table that highlights the weekends, holidays and traditional festivals in China, such as Chinese New Year.

\subsection{Experimental setting}

\begin{table}
  \centering
  \caption{\label{events} Eight promotion events in the forecasting period. Note that Chinese New Year public holidays are not included here since they are not promotion events.}
  \begin{tabular}{llll}
    \toprule
    Name & Description & Start date & End date \\
    \midrule
    11.11 & Singles’ Day promotion & 2020-10-21 & 2020-11-13 \\
    New Year & Chinese New Year's shopping festival & 2021-01-19 & 2021-02-02 \\
    3.8 & Women's Day promotion & 2021-03-04 & 2021-03-08 \\
    Anniversary  & Retailer's anniversary  & 2021-04-19 & 2021-04-25 \\
    6.18 & June 18th promotion & 2021-06-01 & 2021-06-20 \\
    Summer & Summer promotion & 2021-07-11 & 2021-07-18 \\
    8.8 & August 8th promotion & 2021-08-05 & 2021-08-12\\
    9.9 & September 9th promotion & 2021-09-08 & 2021-09-12\\
   \bottomrule
  \end{tabular}
\end{table}

\begin{figure}
  \resizebox{\textwidth}{!}{
    \includegraphics{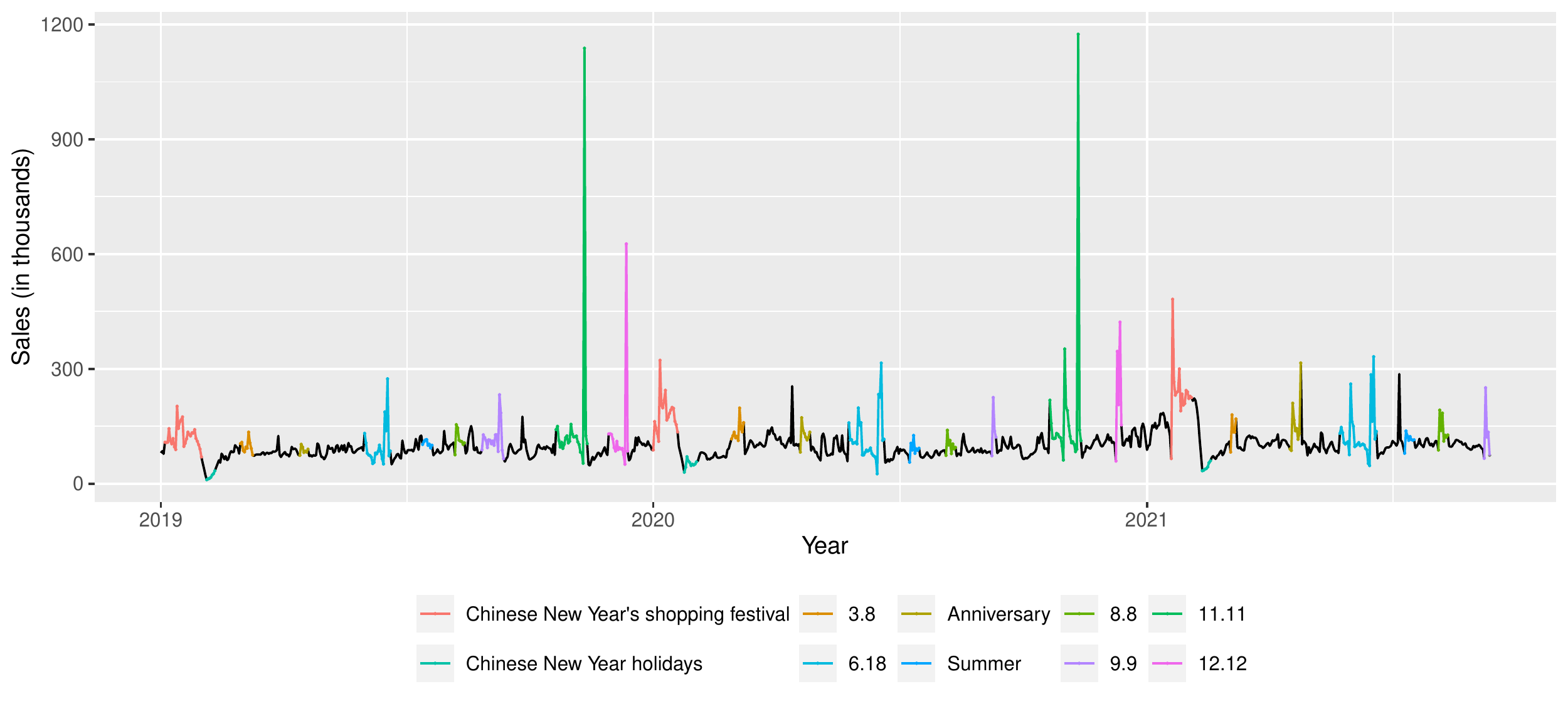}
  }
  \caption{\label{total_sale} Total sales of the category ``food'' from 2019-01-01 to 2021-09-12 in the Chinese online supermarket. }
\end{figure}

Our objective is to forecast sales during eight promotional events from the end of 2020 to 2021-09-12 using observations starting from 2019-01-01. The eight events are listed in Table~\ref{events}. Due to a lack of historical data, we use promotion strength as a predictor rather than using many dummies for each specific type of holiday. Figure~\ref{total_sale} shows a drop in sales during Chinese New Year holidays, e.g., from 2021-02-11 to 2021-02-18. This can be explained by customer stockpiling before the holiday and a slowdown in express service due to staff shortages. Therefore, we also add Chinese New Year holidays as exogenous dummy variables.

We adopt different modelling strategies for different variables. Bottom-level time series containing 60 percent of zeros in the last two months are considered to be intermittent and forecasts are generated using simple exponential smoothing model implemented by \code{ses} function in \pkg{forecast} package for \proglang{R}.
For other non-intermittent series, we first implement Box-Cox transformation, and then base forecasts are generated using a regression model with ARIMA errors. The parameter of Box-Cox transformation is chosen by method in \cite{guerreroTimeseriesAnalysisSupported1993}, while ARIMA orders are chosen using the \code{auto.arima} function in \pkg{forecast} package for \proglang{R}.
The only exception is the top-level series do not implement Box-Cox transformation, for which forecasts without transformation perform better.
When there are only a few training observations in promotion periods, regression models with ARIMA errors can produce unreasonably large forecasts and the Box-Cox transformation can amplify forecasts exponentially. Any forecasts more than ten times larger than the highest sale in history is replaced with the mean of historical observations.

We keep the intermittent series in bottom level immutable to reduce the dimension. We also keep series with more than one year of training observations in bottom level immutable, since we are more confident on their base forecasts obtained from enough historical information. The top-level series is also kept immutable because the signal is strong in the aggregate series.

Sales are non-negative naturally. We conduct two experiments to compare the performance of the method in Section~\ref{sec:method} and compare results both with and without imposing non-negativity.  Due to the different time series lengths in the bottom level, covariance based reconciliation method such as MinT(Shrinkage) cannot easily be used. Furthermore, time series exhibit significantly different patterns in promotion events, and in particular the variance of base forecast errors are higher when conditioning on a promotional period. Therefore for reconciliation, we construct a weighting matrix based on the forecast error variance, computed using historical promotion events only.

\subsection{Forecasting results}

Due to the various scales of different events and time series, we evaluate the forecasts using the average Mean Absolute Scaled Error~\citep[MASE,][]{hyndman2006another}. Table~\ref{sec5_res} presents the mean MASE values of the forecasts of eight promotion events in Table~\ref{events}. We provide average MASE across different types of series. Level 0 and 1 are the top two levels respectively.
The bottom level is split into three. Level 2-0 represents the mutable time series in the bottom level, these have a few training observations especially during promotion events. Level 2-1 refers to the intermittent series in the bottom level which are kept immutable.  Level 2-2 represents the time series that have over one year of training observations, and are also chosen to be immutable. We follow the same taxonomy for the reconciliation methods as described in Section~\ref{sec:simulations}. However here,  ``C'' refers to reconciliation, where in addition to level 0, level 2-1 and level 2-2, are kept immutable. Also, for this application, $\text{WLS}_v$ is based on forecast error variance computed only during promotional periods.

Regarding the weighting scheme, OLS performs the worst at all levels even compared to base forecasts, followed by $\text{WLS}_s$ and $\text{WLS}_v$. Imposing non-negative constraints is can stabilize forecasts at level 2-1 (intermittent series) in particular, especially when using conventional reconciliation. Of greatest interest is the comparison between reconciliation with immutability constraints and without immutability constraints. Even in the best case scenario of $\textrm{WLS}_v$ with nonnegativity imposed, reconciliation without immutability constraints leads to large deterioration in forecasts at Levels 2-1 (intermittent series) and 2-2 (series with more training data) to the point where they are worse than base forecasts. In contrast, the equivalent results where immutability constraints are imposed are at least as good as, or better than base forecasts for all groups of series.

\begin{table}
  \centering
  \caption{\label{sec5_res}Out of sample forecasting accuracies for the Chinese online retailer data.
  Subcolumns ``C'' and ``C+NN'' show forecasting accuracies of immutability reconciliation without and with imposing non-negativity constraints, respectively.
  Subcolumns ``U'' and ``U+NN'' show forecasting accuracies of conventional reconciliation without and with imposing non-negativity constraints, respectively.
  The accuracy of base forecasts is also shown for comparison}.

  \resizebox{\textwidth}{!}{
    \begin{tabular}{lrrrrrrrrrrrrr}
      \toprule
      \multirow{2}{*}{Level} &
      \multirow{2}{*}{Base} &
      \multicolumn{4}{c}{OLS} &
      \multicolumn{4}{c}{$\text{WLS}_s$} &
      \multicolumn{4}{c}{$\text{WLS}_v$} \\
       &  \multicolumn{1}{l}{} &
      \multicolumn{1}{c}{C} &\multicolumn{1}{c}{C+NN} &
      \multicolumn{1}{c}{U} &\multicolumn{1}{c}{U+NN} &
      \multicolumn{1}{c}{C} &\multicolumn{1}{c}{C+NN} &
      \multicolumn{1}{c}{U} &\multicolumn{1}{c}{U+NN} &
      \multicolumn{1}{c}{C} &\multicolumn{1}{c}{C+NN} &
      \multicolumn{1}{c}{U} &\multicolumn{1}{c}{U+NN} \\
      \cmidrule(lr){3-6} \cmidrule(lr){7-10} \cmidrule(lr){11-14}
      0 & 2.94 &  2.94 & 2.94 & 2.93   & 2.92  & 2.94 & 2.94 & 2.72  & 2.72 & 2.94 & 2.94 & 2.75 & 2.77 \\
      1     & 2.66 &  9.31 & 4.94 & 272.83 & 48.84 & 6.41 & 4.83 & 16.09 & 6.50 & 2.43 & 2.47 & 2.39 & 2.40 \\
      2-0   & 2.04 &  8.98 & 4.31 & 3.98   & 2.70  & 7.19 & 3.71 & 2.96  & 2.32 & 1.97 & 1.88 & 1.86 & 1.83 \\
      2-1   & 0.11 &  0.11 & 0.11 & 42.66  & 15.43 & 0.11 & 0.11 & 26.99 & 8.34 & 0.11 & 0.11 & 1.52 & 1.52 \\
      2-2   & 1.08 &  1.08 & 1.08 & 1.64   & 1.48  & 1.08 & 1.08 & 1.36  & 1.25 & 1.08 & 1.08 & 1.58 & 1.19 \\
      \hline
      \end{tabular}}
  \end{table}

\section{A grouped hierarchy: forecasting Wikipedia daily pageviews}
\label{sec:casestudy}

In this section, we illustrate the proposed method using a publicly available real-world dataset - Wikipedia daily pageviews. The dataset consists of one year of daily pageviews (from 2016-06-01 to 2017-06-29) for the articles of the most popular social networks worldwide on Wikipedia~\citep{ashouriTreeBasedMethodsClustering2018}. The dataset has a grouped structure with the following attributes: ``Agent'': Spider, User, ``Access'': Desktop, Mobile app, Mobile web, ``Language'': en (English), de (German), es (Spanish), zh (Chinese) and ``Purpose'': Blogging related, Business, Gaming, General purpose, Life style, Photo sharing, Reunion, Travel, Video (see Table~\ref{emp_structure}).

\begin{table}
  \centering
  \caption{\label{emp_structure}Social networking Wikipedia pageviews grouping structure.}
  \begin{tabular}{llll}
    \hline
    Grouping & Series & Grouping & Series \\
    \hline
    Total & & Language & \\
    & 1. Social Network & &  10. zh(Chinese)\\
    Access & & Purpose & \\
    & 2. Desktop & & 11. Blogging related \\
    & 3. Mobile app & & 12. Business\\
    & 4. Mobile web  & & 13. Gaming\\
    Agent & & & 14. General purpose\\
    & 5. Spider & &  15. Life style\\
    & 6. User & & 16. Photo sharing\\
    Language & & & 17. Reunion\\
    & 7. en (English) & &18. Video\\
    & 8. de (German) & Articles& \\
    & 9. es (Spanish) && ...\\
    \hline
  \end{tabular}
\end{table}

The time series are obtained through the Wikipedia API\footnote{Wikipedia API: \url{https://wikimedia.org/api/rest_v1/}}, and we remove all-zero series. The final dataset contains 1041 time series. We consider the main grouping levels and interactive combination of them for reconciliation, i.e., we use base forecasts of these levels and variance covariance matrix of their base forecast errors to obtain reconciled forecasts. The applied hierarchical levels are as follows: Total, Access, Agent, Language, Purpose, Article, Access $\times$ Agent, Access $\times$ Language, Access $\times$ Purpose, Agent $\times$ Language, Agent $\times$ Purpose, Language $\times$ Purpose and Bottom levels: Access $\times$ Agent $\times $Language $\times$ Purpose $\times$ Article.

Same as Section~\ref{study1},  we keep the top level immutable during reconciliation. We compare the forecasting accuracy of reconciled forecasts between different base forecasting methods, between multiple estimation methods of reconciliation matrix, and between whether equality constraints are included. For simplicity,  the average RMSE of main aggregation levels and their average are reported in Table~\ref{emp_res}.

\begin{table}[]
  \caption{\label{emp_res}Out-of-sample forecasting accuracies of the Wikipedia pageviews dataset.  The upper panel (``ETS'') shows the results using the ETS base forecasts, and the lower panel (``ETS + ARIMA'') shows the results using ETS base forecasts for the top level and ARIMA base forecasts for the other levels. Subcolumns ``U'' and ``C'' show forecast accuracies of unconstrained and constrained forecast reconciliation using different covariance matrices - OLS, $\text{WLS}_s$, $\text{WLS}_v$ and  MinT (Shrinkage). The accuracy of base forecasts is also shown for comparison.}
  \resizebox{\textwidth}{!}{
    \begin{tabular}{lrrrrrrrrr}
      \toprule
      & \multicolumn{9}{c}{ETS} \\ \cmidrule(lr){2-10}
        \multirow{2}{*}{Level} & \multirow{2}{*}{Base} &
          \multicolumn{2}{c}{OLS}                                             & \multicolumn{2}{c}{$\text{WLS}_s$}                                          & \multicolumn{2}{c}{$\text{WLS}_v$}                                          & \multicolumn{2}{c}{MinT (Shrinkage)}                                 \\
          & \multicolumn{1}{l}{}     & \multicolumn{1}{c}{U} & \multicolumn{1}{c}{C} & \multicolumn{1}{c}{U} & \multicolumn{1}{c}{C} & \multicolumn{1}{c}{U} & \multicolumn{1}{c}{C} & \multicolumn{1}{c}{U} & \multicolumn{1}{c}{C} \\ \cmidrule(lr){3-4} \cmidrule(lr){5-6} \cmidrule(lr){7-8} \cmidrule(lr){9-10}

          Total & 12210.156 & 13563.924 & 12210.156 & 15192.699 & 12210.156 & 15433.606 & 12210.156 & 25267.803 & 12210.156\\
Language & 3498.404 & 3888.629 & 3769.815 & 4261.181 & 3823.865 & 4358.279 & 3544.511 & 8135.100 & 4829.509\\
Access & 4980.063 & 5839.776 & 5653.466 & 6353.256 & 5850.465 & 6850.826 & 5622.431 & 14307.679 & 10690.748\\
Agent & 10054.721 & 9714.999 & 9775.191 & 9282.356 & 8834.911 & 9089.362 & 8001.871 & 13717.154 & 9895.496\\
Purpose & 3538.569 & 3422.376 & 3609.756 & 2819.662 & 3052.171 & 2707.560 & 2535.024 & 3635.515 & 2913.612\\
Network & 604.442 & 741.925 & 764.745 & 595.083 & 628.663 & 532.493 & 506.161 & 871.825 & 831.567\\
Bottom & 58.854 & 102.675 & 104.221 & 74.090 & 75.868 & 64.449 & 62.546 & 105.923 & 100.636\\
Average & 4992.173 & 5324.901 & 5126.764 & 5511.190 & 4925.157 & 5576.654 & 4640.386 & 9434.428 & 5924.532\\
\midrule
          & \multicolumn{1}{l}{}     & \multicolumn{1}{l}{}              & \multicolumn{1}{l}{}            & \multicolumn{1}{l}{}              & \multicolumn{1}{l}{}            & \multicolumn{1}{l}{}              & \multicolumn{1}{l}{}            & \multicolumn{1}{l}{}              & \multicolumn{1}{l}{} \\

        \multicolumn{10}{c}{ETS + ARIMA} \\ \cmidrule(lr){2-10}
        \multirow{2}{*}{Level} & \multirow{2}{*}{Base} &
          \multicolumn{2}{c}{OLS}                                             & \multicolumn{2}{c}{$\text{WLS}_s$}                                          & \multicolumn{2}{c}{$\text{WLS}_v$}                                          & \multicolumn{2}{c}{MinT (Shrinkage)}                                 \\
          & \multicolumn{1}{l}{}     & \multicolumn{1}{c}{U} & \multicolumn{1}{c}{C} & \multicolumn{1}{c}{U} & \multicolumn{1}{c}{C} & \multicolumn{1}{c}{U} & \multicolumn{1}{c}{C} & \multicolumn{1}{c}{U} & \multicolumn{1}{c}{C} \\ \cmidrule(lr){3-4} \cmidrule(lr){5-6} \cmidrule(lr){7-8} \cmidrule(lr){9-10}
Total & 12210.156 & 18870.816 & 12210.156 & 24241.874 & 12210.156 & 24707.504 & 12210.156 & 21232.512 & 12210.156\\
Language & 5898.918 & 5615.024 & 6236.067 & 6316.638 & 5124.889 & 6439.814 & 3835.802 & 5634.430 & 3772.354\\
Access & 8511.683 & 8125.442 & 7954.186 & 8953.325 & 6920.617 & 9134.150 & 4942.954 & 7525.282 & 5558.481\\
Agent & 12831.442 & 11617.428 & 11757.789 & 12598.884 & 10619.791 & 12663.466 & 7775.078 & 11004.191 & 7525.498\\
Purpose & 4090.289 & 4295.386 & 5005.950 & 3403.371 & 3384.287 & 3251.668 & 2340.395 & 3137.897 & 2629.750\\
Network & 563.297 & 653.120 & 749.176 & 521.706 & 613.466 & 479.970 & 425.459 & 462.020 & 424.921\\
Bottom & 55.010 & 88.066 & 99.331 & 66.444 & 73.399 & 54.911 & 56.332 & 55.092 & 56.730\\
Average & 7350.964 & 8196.203 & 7318.887 & 9339.300 & 6478.868 & 9446.095 & 5254.974 & 8166.055 & 5353.527\\

 \bottomrule

          & \multicolumn{1}{l}{}     & \multicolumn{1}{l}{}              & \multicolumn{1}{l}{}            & \multicolumn{1}{l}{}              & \multicolumn{1}{l}{}            & \multicolumn{1}{l}{}              & \multicolumn{1}{l}{}            & \multicolumn{1}{l}{}              & \multicolumn{1}{l}{}
      \end{tabular}}
  \end{table}

The upper panel of Table~\ref{emp_res} shows the results of using ETS base forecasts for all levels. We can see that the constrained forecast reconciliation with $\text{WLS}_s$ covariance matrix improves forecasting accuracy in upper levels (i.e., levels upon the Purpose level) but decreases accuracy in other levels. The constrained forecast reconciliation with $\text{WLS}_v$ and MinT (Shrinkage) improves forecasting accuracy across all levels compared to the unconstrained forecast reconciliation.

The lower panel shows the results of using ETS to generate base forecasts for the top level and ARIMA for other levels. ETS base forecasts are more accurate compared to ARIMA base forecasts in most levels except the bottom level, as shown in the Base column of Table~\ref{emp_res}. The difference between constrained and unconstrained forecast reconciliation in the OLS and WLS$_v$ columns is similar to that in the upper panel, but the forecasting performances are worse. The constrained forecast reconciliation using WLS$_v$ and MinT (Shrinkage) covariance matrix shows more significant improvements in accuracy except for the tiny decreases in bottom levels. The results in both panels show that keeping base forecasts of the top level immutable can capture the diversity and dependency among different levels more wisely, and engage the superiority of forecast reconciliation when covariance matrix of base forecast errors is used and the time series in disaggregated levels are noisy.

\section{Conclusions}
\label{sec:conclusions}

In this paper, we propose a forecast reconciliation approach that can keep the base forecasts of specific levels or multiple nodes from different levels immutable after reconciliation. The proposed method is flexible and general enough to allow for expert judgment in choosing the series with forecasts that should not be adjusted by reconciliation. We prove that the proposed method can produce unbiased reconciled forecasts as long as the base forecasts are unbiased, and the equality constraints do not go beyond the boundary conditions.

Monte Carlo simulations and two empirical applications show the superiority of the proposed method over conventional forecast reconciliation. In particular, constrained forecast reconciliation shows promising results when series are noisy and decision-makers have limited knowledge about the underlying data generating processes of lower levels in the hierarchy. The application to sales data from a major Chinese online retailer, shows the potential of the proposed method in reconciling the forecasts of high-dimensional hierarchy where careful judgment is used in selecting immutable time series.

There are several valuable directions worthy of further investigation. First, we show the results under the cross-sectional scenario and propose the framework is suitable for both cross-sectional and temporal hierarchies. As such, the framework can also be extended into the cross-temporal data as in \cite{kourentzesCrosstemporalCoherentForecasts2019}. Second, we select the set of immutable series using judgment. It is worth to explore whether theoretical properties of times series can be used to automate this process. Due to the diversity between different forecasting methods and the heterogeneity across different datasets, the performance of the forecast reconciliation framework may not be robust, therefore further research, particularly on impact of the covariance matrix estimation is needed.

\section*{Acknowledgments}

Yanfei Kang is supported by the National Natural Science Foundation of China
(No. 72171011). Feng Li is supported by the Beijing Universities Advanced Disciplines
Initiative (No. GJJ2019163) and the Emerging Interdisciplinary Project of CUFE. This research was supported by Alibaba Group through the Alibaba Innovative Research Program and the high-performance computing (HPC) resources at Beihang University.

\appendix

\section{Proof for unbiased reconciled forecasts}
\label{appendixA}
\cite{panagiotelisForecastReconciliationGeometric2021} show that the unbiased preserving property of forecast reconciliation will hold as long as $\bm{SG}$ is a projection onto the space spanned by the columns of $\bm{S}$. Algebraically, this can be proven by showing $\bm{G}\bm{S}=\bm{I}$ for a given choice of $\bm{G}$. For the choice $\hat{\bm{G}}$ shown in Equation~\ref{eq:optgim} in Section~\ref{sec:method} we can show,

\[
\begin{aligned}
\hat{\boldsymbol{G}}\boldsymbol{S} &= \left[\begin{matrix*}[l]
  \hat{\boldsymbol{G}}_1 & \boldsymbol{0}_{(m-k)\times k} \\
  \boldsymbol{0}_{k \times (n-k)} & \boldsymbol{I}_{k}
\end{matrix*}\right](\boldsymbol{I}_n - \left[\begin{matrix*}[l]
  \boldsymbol{0}_{(n-m)\times (n-k)} & \boldsymbol{S}_2 \\
  \boldsymbol{0}_{m\times (n-k)} & \boldsymbol{0}_{m\times k}
\end{matrix*}\right]) \left[\begin{matrix*}[l]
  \boldsymbol{S}_1 & \boldsymbol{S}_2 \\
  \boldsymbol{I}_{m-k} & \boldsymbol{0}_{(m-k)\times k} \\
  \boldsymbol{0}_{k \times (m-k)} & \boldsymbol{I}_{k\times k}
\end{matrix*}\right] \\
&= \left[\begin{matrix*}[l]
\boldsymbol{I}_{m-k} & \hat{\boldsymbol{G}_1}\check{\boldsymbol{S}}_2 \\ \boldsymbol{0}_{k\times (m-k)} & \boldsymbol{I}_k
\end{matrix*}\right] -
\left[
\begin{matrix*}[l]
\boldsymbol{0}_{m-k} & \hat{\boldsymbol{G}}_1 \check{\boldsymbol{S}}_2 \\ \boldsymbol{0}_{k\times (m-k)} & \boldsymbol{0}_{k}
\end{matrix*}
\right] \\
&= \boldsymbol{I}_m,
\end{aligned}
\]
where
\[
\check{\boldsymbol{S}}_2 = \left[\begin{matrix*}[l]
  \boldsymbol{S}_2 \\ \bm{0}_{(m-k)\times k}
\end{matrix*}\right].
\]

\section{Proof for validity of candidate basis sets}
\label{appendixB}

A candidate basis set $\bm{b}^*$ is valid if $\bm{y}$ can be constructed when only $\bm{b}^*$ and the hierarchy are known.
By construction,
\[
\bm{b}^* =   \bm{S}_{\{j\}}\bm{b},
\]
To obtain a representation of $\bm{y}$ in terms of $\bm{b^*}$, we can use
\[\bm{y} = \bm{Sb} = \bm{SS}_{\{j\}}^{-1}\bm{b}^* = \bm{S}^*\bm{b}^*,\]
where $\bm{S}^* = \bm{SS}_{\{j\}}^{-1}$, which can be obtained only when $\bm{S}_{j}$ is invertible.


\newpage

\bibliographystyle{agsm}
\bibliography{references.bib}

\end{document}